\documentclass[manuscript]{aastex}

\shorttitle{Remote Globular Clusters in M33}
\shortauthors{Huxor et al.}

\begin{document}


\title{The Discovery of Remote Globular Clusters in M33}


\author{A. Huxor\altaffilmark{1},  A. M. N. Ferguson, M. K. Barker}
\affil{Institute for Astronomy, University of Edinburgh, Royal Observatory, Blackford Hill, Edinburgh UK EH9 3HJ, UK}

\author{N. R. Tanvir}
\affil{Department of Physics and Astronomy, University of Leicester, University Road, Leicester UK LE1 7RH, UK}

\author{M. J. Irwin, S. C. Chapman} 
\affil{Institute of Astronomy, Madingley Road, Cambridge, UK CB3 0HA, UK}

\author{R. Ibata}
\affil{Observatoire de Strasbourg, 11, rue de l'Universite, F-67000, Strasbourg, France}

\and

\author{G. Lewis}
\affil{Sydney Institute for Astronomy, School of Physics, A29, University of Sydney, NSW 2006, Australia}


\altaffiltext{1}{present address: H.H.Wills Physics Laboratory, Tyndall Avenue, Bristol, BS8 1TL, UK }


\begin{abstract}

We present the discovery of four remote star clusters in M33, one of
which is of an extended nature. Three of the clusters were discovered
using survey data from the Isaac Newton Telescope Wide-Field Camera
while one was discovered serendipitously in a deep image taken with
the Hubble Space Telescope's Advanced Camera for Surveys.  With
projected radii of 38--113 arcmin (9.6--28.5 kpc for an assumed M33
distance of 870~kpc), these clusters lie significantly beyond all but
one of the currently-confirmed clusters in M33.  The clusters have
magnitudes and colors consistent with their being old to
intermediate-age globular clusters.  Indeed, they bear a strong
resemblance to the outer halo GC population of the Milky Way and M31
in terms (V-I)$_0$ colours.  The three outermost clusters are
projected on the far side of M33 with respect to M31, an asymmetry
that could suggest tidal interactions have affected M33's globular
cluster distribution at large radii.
\end{abstract}


\keywords{galaxies: globular clusters: -- galaxies: individual (M33)
-- galaxies: formation -- galaxies: evolution -- galaxies: star
clusters -- galaxies: stellar content }



\section{Introduction}

Globular clusters (GCs) have proved to be valuable probes of galaxy
formation, providing key clues about the assembly histories of the
Milky Way (hereafter MW) (e.g. \citealt{SearleZinn78}), and M31
(e.g. \citealt{Perrett02}). A particular property of GCs that makes them
especially important is that they can often be found out to extremely
large radii in galaxy halos.  Such outer halo GCs are key tracers of
the baryonic content and dynamics of galaxies on scales of tens to
hundreds of kiloparsecs (e.g. \citealt{EvansWilkinson00}).

First discovered by \citet{MadoreArp79}, AM-1 remains the most distant
GC known in the MW with a galactocentric distance of 123~kpc.  M31
hosts GCs at much larger radial distances, out to $\gtrsim 200$~kpc,
however this population has only been uncovered in the last few years
thanks to sensitive panoramic digital imaging surveys
(e.g. \citealt{Martinetal06, Huxoretal08}).  For example, using 84
square degrees of imaging data from the Isaac Newton Telescope
Wide-Field Camera (INT/WFC) and Canada France Hawaii Telescope (CFHT)
MegaCam surveys of M31, \citet{Huxoretal08} were able to increase the
number of known GCs with projected radii $\gtrsim 1$ degree ($\approx
14$~kpc at M31's distance) by more than 75\%; this is especially
remarkable given that most of this data came from a single outer
quadrant of M31 and suggests that many more halo GCs remain to be
discovered in that system.

M33 is the least massive spiral in the Local Group and there has been
much interest in understanding how its star cluster system compares to
that of its larger brethren. Previous studies have suggested that,
like the MW and M31, M33 has a population of star clusters which
exhibit halo kinematics but that these clusters possess a much larger
age spread and a higher mean metallicity than their MW counterparts
(e.g.  \citealt{Sarajedini00, Chandaretal02}). Indeed, M33 appears to
have a cluster population with a very broad range of ages, bearing a
stronger resemblance to the cluster populations of the lower mass
Magellanic Clouds rather than the MW or M31.

One drawback of all M33 star cluster studies to date is that they have
focused only on objects projected on the inner regions of the galaxy,
with little exploration of the galaxy outskirts. This bias has made it
rather difficult to unambiguously associate clusters with the disk or
the halo. The \citet{Chandaretal02} study, for example, includes only
objects that lie within a projected radius of 5~kpc, corresponding to
just over half the isophotal R$_{25}$ radius of 9~kpc
\citep{deVauc91}. \citet[][hereafter SM07] {SarajediniMancone07} have
recently compiled a list of all M33 star clusters and cluster
candidates identified in all previous studies.  This has led to the
creation of a unified catalog of $\sim 300$ $`$high confidence'
objects. The most
remote cluster reported in the original SM07 catalog lies at a mere
7~kpc in projected radius. Since the publication of SM07,
\citet[][hereafter ZKH08]{Zloczewskietal08} have conducted a search
for new star clusters in the outskirts of M33 based on 0.75 square
degrees of deep CFHT Megacam imagery.  Although their survey has led
to the discovery of more than 120 new star cluster candidates, none of
these lie beyond a projected galactocentric radius of 8.3~kpc.

In order to establish whether M33, like its larger neighbours, also
possesses a population of remote halo star clusters, we have
undertaken a search for GCs in INT/WFC survey data of M33, as well as
in our own deep Hubble Space Telescope Advanced Camera for Surveys
(HST/ACS) imagery of two outer fields in M33. Our search has resulted
in the identification of four new outlying star clusters in M33, which
lie at projected radii of 10--30~kpc. A fifth cluster was also
uncovered in our search but it has since been published by
\citet{Stonkuteetal08} who independently found the object in Subaru
Suprime-Cam data (M33-EC1).  Throughout this paper, we adopt a
distance modulus to M33 of $(m-M)_{0}=24.69$, corresponding to
870~kpc, for consistency with SM07.

\section{Observations}

%% In a manner similar to \objectname authors can provide links to dataset
%% hosted at participating data centers via the \dataset{} command.  The
%% second curly bracket argument is printed in the text while the first
%% parentheses argument serves as the valid data set identifier.  Large
%% lists of data set are best provided in a table (see Table 3 for an example).
%% Valid data set identifiers should be obtained from the data center that
%% is currently hosting the data.
%%
%% Note that AASTeX interprets everything between the curly braces in the 
%% macro as regular text, so any special characters, e.g. "#" or "_," must be 
%% preceded by a backslash. Otherwise, you will get a LaTeX error when you 
%% compile your manuscript.  Special characters do not 
%% need to be escaped in the optional, square-bracket argument.

\begin{figure}
\epsscale{.80}
\plotone{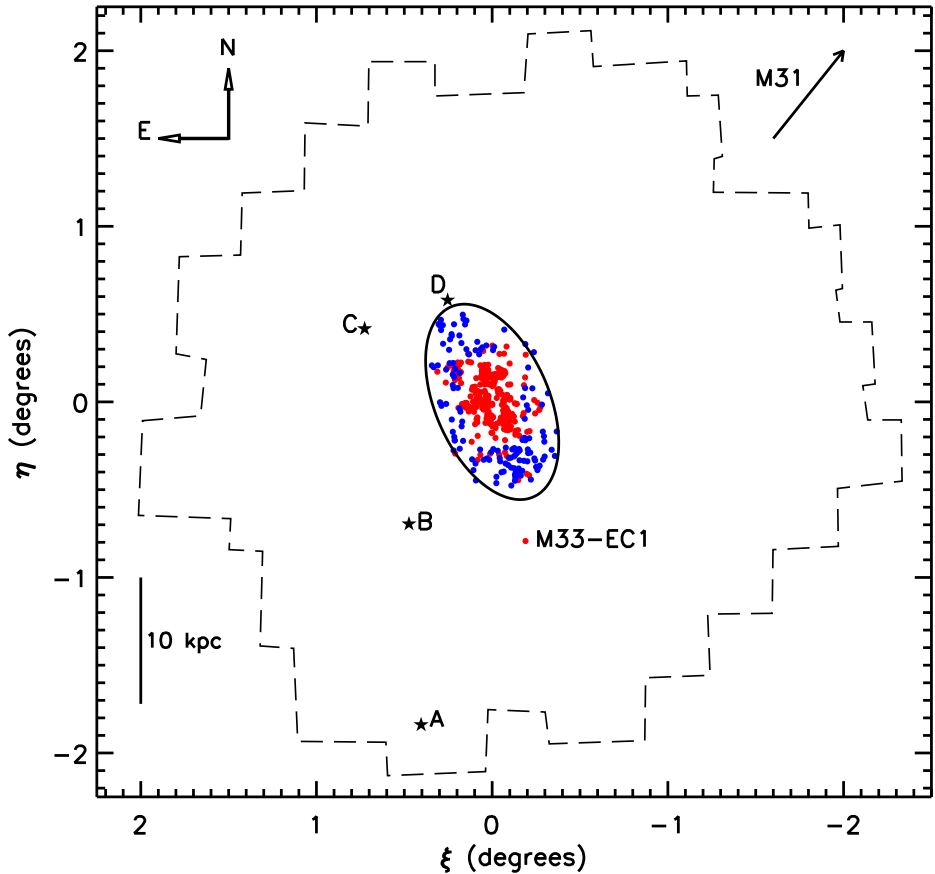}
\caption{The locations of the new clusters (black stars) are shown in relation to the $`$high confidence'
clusters in the SM07 catalogue (red circles) and the $`$probable
clusters' of ZKH08 (blue circles). The \citet{Stonkuteetal08} cluster,
M33-EC1, is labeled, and an arrow indicates the direction to M31.  The
solid ellipse indicates the R$_{25}$ isophote and the dashed outline
indicates the extent of our INT/WFC survey coverage. The outermost of
the new clusters, along with the remote extended cluster of
\citet{Stonkuteetal08}, project on the far side of M33 with respect to M31.
\label{Fi:figLocations}}
\end{figure}

The primary dataset used in our GC search is the INT/WFC survey of M33
which was conducted over several observing runs in 2002-2008. The WFC
is a four-chip CCD mosaic camera with a 0.29 deg$^{2}$ field of view
and a pixel scale of 0.33 arcsec pixel$^{-1}$, equipped to the INT
2.5m telescope on La Palma. Exposures of 900-1200 seconds were taken
in both the Johnson V and Gunn i bands, with a median seeing of
1.2\arcsec.  This depth is sufficient to detect individual stars in
M33 to V$\sim24.5$ and i$\sim~23.5$ with a typical signal-to-noise of
5.  The fields observed comprise a contiguous, almost circular,
region, covering approximately 12 square degrees, very much greater in
extent than any previously-published survey of M33 (see Figure
\ref{Fi:figLocations}).  The survey data were processed using the
standard INT Wide Field Survey pipeline provided by the Cambridge
Astronomical Survey Unit; as discussed in \citet{IrwinLewis01}, this
pipeline provides basic data processing and astrometric and
photometric calibration, as well as object detection and
classification.

Our star cluster search concentrated on only those regions beyond the
main optical disk of M31. We adopted a strategy similar to that used
in our search for GCs around M31 \citep{Huxoretal08} and which
exploits the fact that clusters at the distance of M31 and M33 should
be just resolved in good seeing. Indeed, given the survey depth, we
expect detection of individual red giants down to $\sim3$ magnitudes
below the tip of the red giant branch. Initial GC candidate selection
was based on source magnitude, color and morphological classification
(stellar/non-stellar) provided by the photometric pipeline, exploiting
the fact that most known GCs lie within specific ranges of absolute
magnitude ($-10.5 \lesssim M_V \lesssim -3.5$, equivalent to $14
\lesssim V \lesssim 21$ at the distance of M33) and color ($0 \lesssim
V-I \lesssim 1.7$).  These generous regions of parameter space are
sufficient to allow for photometric errors and avoid exclusion of any
slightly atypical GCs from the candidate list generated. Unlike our
search for M31 GCs, ellipticity and object FWHM were not employed as
search criteria since the number of initial candidates selected using
just magnitude, color and morphology alone was not prohibitively large
and all candidates could be visually inspected.  As in
\citet{Huxoretal08}, every single survey image was also visually
inspected in order to identify well-resolved diffuse extended
GCs. Such objects are rather common in the outskirts of M31
(e.g. \citealt{Huxoretal05, Huxoretal08}) and would not have been
automatically found by the method above as they do not appear as a
single source in the object catalog.
                                               
We also conducted a search for outlying star clusters in two deep
HST/ACS pointings that we obtained along M33's northern major axis at
radii of 36$\arcmin$ and 46$\arcmin$. These data were taken in the
F606W and F814W filters as part of GO PID 9837 (PI = A. Ferguson) with
total exposure times of 18050 and 12955 secs respectively and are
being used for an analysis of the star formation history of M33's
outer disk (Barker et al., in preparation).  The star cluster search
in this dataset was conducted purely by visual inspection of the
drizzled stacked frames since star clusters at the distance of M33 are
very well-resolved in HST/ACS imagery.
                                                                              
%% In this section, we use  the \subsection command to set off
%% a subsection.  \footnote is used to insert a footnote to the text. 


%% This section also includes several of the displayed math environments
%% mentioned in the Author Guide.

\section{Results}

Our search resulted in the identification of four new outlying star
clusters in M33, three of which were discovered in the INT/WFC survey
data and one in the HST/ACS data. It is worth noting that although the
HST/ACS frames lie within the INT/WFC areal coverage, the HST cluster
was not independently found in the INT/WFC survey data.  This cluster
is faint (see below) but, more importantly, rather compact in
appearance hence it was not possible to distinguish it as a cluster in
our ground-based imaging.  Figure \ref{Fi:figLocations} shows the
locations of these objects with respect to the 'high confidence'
clusters from SM07 and the ZHK candidate clusters. Our new discoveries
lie at galactocentric radii ranging from 38--113 arcmin, which
corresponds to 9.6--28.5~kpc for an assumed M33 distance of 870~kpc,
and are clearly more remote than all but one of the known population.
If these clusters are associated with M33's disk rather than halo,
then their deprojected radii would be even more extreme, corresponding
to 10--38~kpc.

\begin{figure}
\epsscale{.80}
\plotone{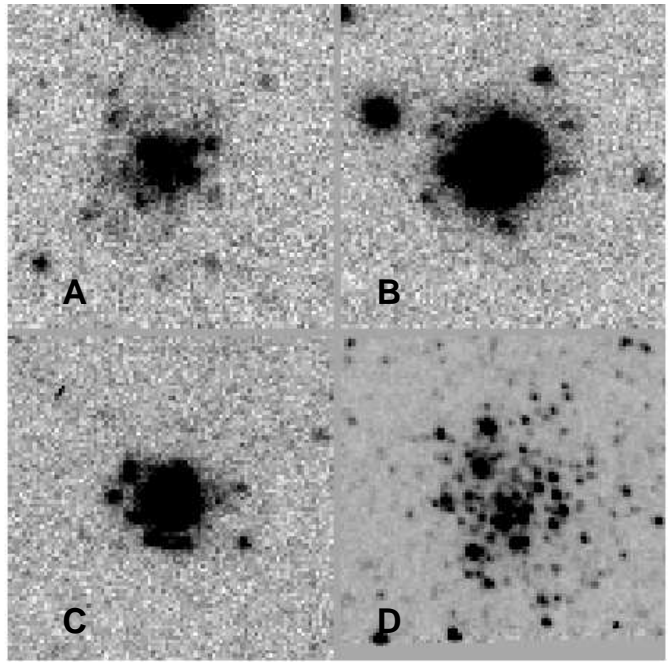}
\caption{Thumbnails (i-band) of the outer halo clusters found in the INT-WFC survey (A, B and C)
and in the F606W HST/ACS data (D).  The INT images span $30\arcsec
\times 30 \arcsec$ while the ACS image spans $7\arcsec \times
7\arcsec$. North is up and east is to the left.  \label{figMosaic}}
\end{figure}

Figure \ref{figMosaic} shows images of the four newly discovered
clusters. Two of these clusters (HM33-B and HM33-C) are high surface
brightness and compact while one cluster (HM33-A) is diffuse and
extended, much like M33-EC1 reported by \citet{Stonkuteetal08}.
Figure 2 also shows the stacked F606W image of the cluster (HM33-D)
discovered in the HST/ACS image.

Integrated photometry of the new INT/WFC clusters was undertaken with
the IRAF task {\it apphot} \footnote{IRAF is distributed by the
National Optical Astronomy Observatories, which are operated by the
Association of Universities for Research in Astronomy, Inc., under
cooperative agreement with the National Science Foundation.}.  Fluxes
within an aperture of radius $8\arcsec$ were obtained for clusters
HM33-B and HM33-C; this aperture was selected to allow direct
comparison with our photometry of M31 GCs \citet{Huxoretal08}.  The
(V-I) color for these clusters was measured within a smaller aperture
of $4\arcsec$ radius that as chosen to reduce the error from the
background and assumes that there is no color gradient in the
clusters. For the extended cluster HM33-A, the magnitudes were
determined within a larger aperture of 12\arcsec\ radius so as to
enclose the bulk of the light. The INT/WFC survey uses Johnson V
(V$^{\prime}$) and Gunn i passbands hence a color transformation was
required to obtain the Johnson/Cousins equivalents V and I.  Following
\citet{McConnachieetal03}, we adopted $V = V^{\prime} + 0.005(V-I)$ and
$I = i -0.101(V-I)$.  Due to the high luminosity of the clusters, the
formal errors reported by {\it apphot} are quite small. The cluster
magnitude uncertainties are dominated by the zero-point errors of
$\sim$0.02 mag while colour errors are estimated to be $\sim$ 0.03
mag.  These errors do not account for contamination from foreground
stars or background galaxies as these are difficult to estimate.
Finally, the cluster magnitudes and colors were corrected for
extinction by interpolating within the \citet{Schlegeletal98}
maps\footnote{http://astro.berkeley.edu/$\sim$marc/dust/data/data.html}
to the position of each cluster.  The INT/WFC clusters have absolute
magnitudes in the range M$_{V_{0}} \sim -6$ to $-7$ placing them near
the peak of the GC luminosity function.

For the HST/ACS cluster, we followed a similar procedure to that of
\citet{Sarajedinietal06}.  First, the F606W drizzled image was
convolved with a Gaussian kernel (1 $\sigma$ = 20 pixels) as otherwise
the light distribution was too patchy to determine the cluster centre
by eye or by fitting a Gaussian.  Once the center was determined,
magnitudes were measured on two of the drizzled sub-images produced by
the STScI CALACS pipeline, with total exposures of 5240 and 10480
seconds for F606W and F814W filters respectively.  We determine fluxes
in a circular aperture of radius 3.5"; as the cluster lies near the
edge of an ACS chip, this was the maximum aperture size that could be
chosen which lay entirely on the chip.  Interpolating the
\citet{Schlegeletal98} dust maps to the cluster's position gave a
reddening of E(B-V) = 0.046.  If we adopt the Sirianni et al. (2005)
extinction ratios for a G2 star in the ACS/WFC filter system and the
most up-to-date ACS
zeropoints\footnote{http://www.stsci.edu/hst/acs/analysis/zeropoints.}
then the dereddened magnitude and color are F606W$_{0}$ = 20.46 and
(F606W-F814W)$_{0}$ = 0.77. The S/N of these images is very high and
the formal uncertainties on the magnitudes are very small. To convert
the ACS/WFC magnitudes to the Johnson-Cousins system, we used the
synthetic transformations of Sirianni et al. (2005) which gave V$_{0}$
= 20.60 (with a systematic zero-point error of 0.05 mag) and
(V-I)$_{0}$ = 0.92 mags.  This corresponds to a cluster magnitude of
M$_{V_{0}}$ of -4.09, the faintest yet known star cluster in M33, and
is comparable to the low-luminosity MW cluster Pal 13.  Table
\ref{Tab:tblNewGCs} lists the positional data and corrected photometry
for all clusters.

\section{Discussion}

In the following, the newly-discovered clusters are compared with the
known cluster population in M33. For this purpose, we use a subset of
167 of the 296 $`$high confidence' clusters in the SM07 catalogue that
have both V and I magnitudes available.  We corrected this sample for
extinction by interpolating within the \citet{Schlegeletal98} dust
maps to the position of each cluster. This resulted in a range of
A$_V$ from 0.14 to 0.66.  We do not use the ZKH08 sample in these
comparisons as the authors note that the new GCs in their catalog are
only ``probable", and may contain contaminants.

\begin{figure}
\epsscale{.80}
\includegraphics[angle=90,width=15cm]{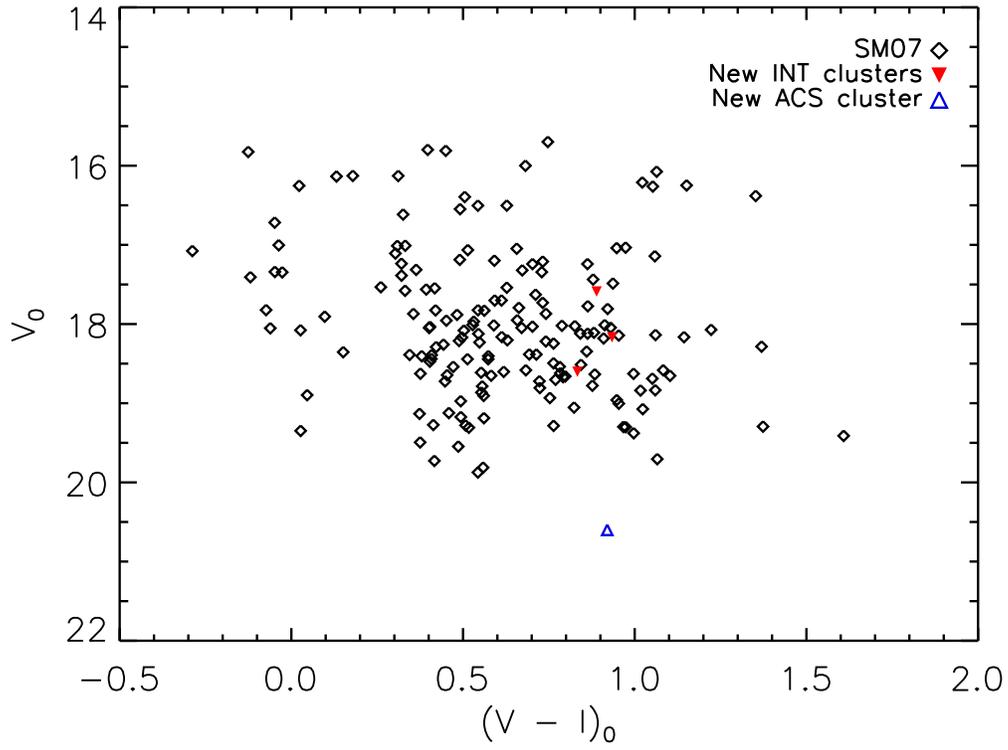}
\caption{Plot of extinction-corrected color versus apparent magnitude for the 
new INT/WFC clusters (inverted triangles), the new ACS cluster
(triangle) and the $`$high-confidence' SM07 sample (open diamonds).\label{Fi:ColMag}}
\end{figure}

Figure \ref{Fi:ColMag} shows the location of the new star clusters on
a plot of V$_0$ magnitude versus (V$-$I)$_0$ color.  There is a broad
spread in magnitudes at all colors and the new objects sit comfortably
within this distribution. On the other hand, the HST/ACS cluster is
far fainter than any previously-confirmed M33 star cluster. However,
the color of HM33-D, combined with its compact appearance on the
HST/ACS images, supports its identification as an intermediate or old
age GC. This conclusion is further strengthened by the isochrone fits
to the color-magnitude diagram which indicate an age of $\gtrsim
3$~Gyr and [M/H]$\gtrsim -0.7$ dex (Barker et al., in preparation).
The existence of such a cluster supports the notion that a population
of very faint GCs may await discovery in deeper panoramic imaging of
the outskirts of M33.

\begin{figure}
\epsscale{.80}
\includegraphics[angle=90,width=15cm]{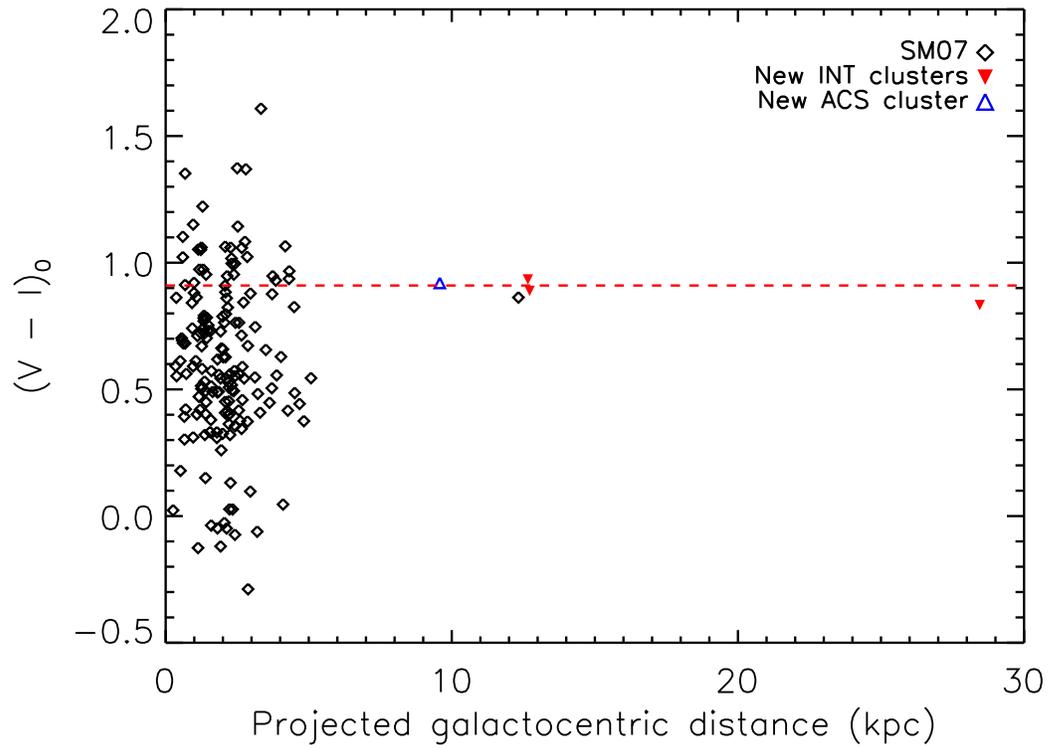}
\caption{Plot of  the extinction-corrected color versus radius for the 
cluster population in M33. The dashed line indicates the mean
extinction-corrected color of the outer halo globular clusters in the
MW and M31 from Huxor et al. (2009).\label{Fi:figColDist}}
\end{figure}

Our newly-discovered clusters, along with that of
\citet{Stonkuteetal08}, show a notable lack of spread in (V$-$I)$_0$
color when compared to the remaining M33 cluster population. This is
particularly apparent when (V$-$I)$_0$ color is plotted against
projected radial distance from M33 (Figure \ref{Fi:figColDist}).  The
new clusters are slightly redder than the overall cluster population
having an average (V$-$I)$_{0}$ color of $ 0.88\pm0.05$ compared to
the average (V-I)$_{0}$ for the inner population being 0.67$\pm0.30$
mags.  The large spread in color at smaller radii is not surprising
since the SM07 sample contains many young clusters, which have bluer
colors than typical old GCs, and reddening issues will be more
problematic here. The colors of the five outermost clusters in Figure
4 are very similar to those of M31 halo GCs with projected radii
greater than $30$~kpc, which have a mean value for (V-I)$_{0}=0.91\pm
0.15$, as well to Milky Way clusters beyond 15~kpc,
which have (V-I)$_{0}=0.92\pm 0.12$ (Huxor et al. 2009). That the MW and
M31 outer halo GCs are generally old ($\gtrsim 10$~Gyr) and metal-poor
([Fe/H]$\lesssim -1.5$)
\citep{harris96, Mackeyetal06, Mackeyetal07} supports the idea that the new M33
halo clusters are archetypical GCs too, albeit with cluster HM33-A and
M33-EC1 having more extended structures.  Indeed,
\citet{Stonkuteetal08} were able to place some constraints on the age
and metallicity of M33-EC1 from analysis of their color-magnitude
diagram, finding an age $\gtrsim 7$~Gyr and [M/H]$\lesssim -1.4$.

Finally, it is notable that so few luminous objects have been
discovered in the $\sim 12$ square degree area we have surveyed around
M33 (Figure \ref{Fi:figLocations}).  Ignoring the central square
degree of our survey which was not searched for new objects due to the
presence of the bright disk, we calculate a GC surface density of
$\approx 0.4$~deg$^{-2}$ in the outskirts of M33 which is almost half
that derived for the GC surface density in M31 over the radial range
$30-100$~kpc (Huxor et al. 2009).  Given the extent of the M33 survey
area, we can be confident that we have found most of the luminous GCs
in the outer halo of the galaxy, although it is possible that a
population of small, compact and faint clusters, such as the example
found in our HST/ACS imagery, still remains to be uncovered. It is
also rather interesting that the outermost clusters in M33 are all
projected on the far side of the galaxy with respect to M31, despite the
survey coverage being essentially uniform.  Assuming the outermost
clusters are moving on circular orbits with velocities of
80~km~s$^{-1}$, typical of the old halo population identified in the
inner regions of M33 (Chandar et al. 2002), their orbital times will
be $\sim 1-2 \times 10^9$~yr raising the question as to whether their
present spatial configuration can be long-lived or just a chance
coincidence.  The clear asymmetry in the outer cluster distribution
could be a signature that tidal stripping by its larger neighbour may
have dramatically affected the globular cluster distribution in M33.
On the other hand, this may just be a chance alignment.  Line of sight
distances, radial velocities and sensitive cluster searches at larger
radii will be required to test these ideas further.

\section{Summary}

In this paper we have presented the discovery of four new outlying
star clusters in M33 using INT/WFC survey data and deep HST/ACS
imagery.  Three of these clusters are compact while one is of the
extended class. These objects, which have projected radii of 38--113
arcmin or 9.6--28.5 kpc (for an assumed M33 distance of 870~kpc), lie
significantly beyond all but one of the currently-confirmed star
clusters in M33. Their colors and magnitudes are typical of the
intermediate-age to old M33 GC population, as well as old age GCs in
the far outskirts of the MW and M31.  The surface density of halo GCs
beyond the optical disk of M33 is roughly one half of that measured in
M31 in the radial range $\sim 30-100$~kpc.  The new clusters have a
strikingly asymmetric distribution around M33, all lying on the far
side of the galaxy with respect to M31.  Further observations are
required to properly characterize the stellar populations of these new
objects and determine their line-of-sight distances and kinematics.

\acknowledgments

APH, AMNF and MKB are supported by a Marie Curie Excellence Grant from
the European Commission under contract MCEXT-CT-2005-025869. NRT
acknowledges a STFC Senior Research Fellowship. We thank Dougal Mackey
for useful discussions, and the anonymous referee whose comments 
greatly improved the paper. The Isaac Newton Telescope is operated on the
island of La Palma by the Isaac Newton Group in the Spanish
Observatorio del Roque de los Muchachos of the Instituto de
Astrofisica de Canarias. This work is partly based on observations
made with the NASA/ESA Hubble Space Telescope obtained at the Space
Telescope Science Institute, which is operated by the Association of
Universities for Research in Astronomy, Inc., under NASA contract NAS
5-26555. These observations are associated with program \#GO9837.


%% in parentheses, after the keyword, but they will not be verified.

{\it Facilities:} \facility{INT(WFC)}, \facility{HST(ACS)}.

\clearpage

\clearpage

\begin{deluxetable}{llrccccc}
\tabletypesize{\scriptsize}
\tablecaption{Basic Data for the New Outlying GCs \label{Tab:tblNewGCs}}
\tablewidth{0pt}
\tablehead{
\colhead{ID} & \colhead{RA} & \colhead{Dec} & \colhead{V$_{0}$} & 
\colhead{(V-I)$_{0}$} & \colhead{A$_{V}$} & \colhead{Projected Galactocentric} & \colhead{Deprojected Galactocentric} \\
\colhead{} & \colhead{J2000} & \colhead{J2000} & \colhead{mags} & 
\colhead{mags} &  \colhead{mags} & \colhead{Radius (kpc)} & \colhead{Radius (kpc)\tablenotemark{a}} 
}
\startdata
HM33-A  	& 01 35 41.78 & +28 49 15.5 & 18.60 &   0.83  & 0.26 & 28.6 & 37.7 \\  %EGC2(FF2)
HM33-B     	& 01 36 02.12 & +29 57 49.4 & 17.59 &  0.89 & 0.17 & 12.8 & 20.4 \\  % GC1(B)
HM33-C  	& 01 37 14.46 & +31 04 27.8 & 18.17 & 0.93 & 0.18 & 12.7  & 17.1 \\  %GC2(A)
HM33-D 		& 01 35 02.20 & +31 14 21.3 & 20.60 & 0.92 & 0.13\tablenotemark{b} & 9.6 & 9.6 \\
\enddata
%% Text for table notes should follow after the \enddata but before
%% the \end{deluxetable}. Make sure there is at least one \tablenotemark
%% in the table for each \tablenotetext.
\tablenotetext{a}{Deprojected values calculated assuming $i=56$ degrees and $PA=23$ degrees.}
\tablenotetext{b}{Extinction in F606W.}
\end{deluxetable}

%% If you use the table environment, please indicate horizontal rules using
%% Do not put multiple tabular environments within a single table.
%% The optional \label should appear inside the \caption command.

\clearpage

%% the file, uncomment the \plotone, \plottwo, and
%% If you need a layout that cannot be achieved with \plotone or
%% \plottwo, you can invoke the graphicx package directly with the
%% \includegraphics command or use \plotfiddle. For more information,
%% please see the tutorial on "Using Electronic Art with AASTeX" in the
%% documentation section at the AASTeX Web site,
%% http://www.journals.uchicago.edu/AAS/AASTeX.
%%
%% The examples below also include sample markup for submission of
%% supplemental electronic materials. As always, be sure to check
%% the instructions to authors for the journal you are submitting to
%% for specific submissions guidelines as they vary from

%% This example uses \plotone to include an EPS file scaled to
%% 80% of its natural size with \epsscale. Its caption
%% has been written to indicate that additional figure parts will be
%% available in the electronic journal.

%% Tables may also be prepared as separate files. See the accompanying
%% sample file table.tex for an example of an external table file.
%% To include an external file in your main document, use the \input
%% command. Uncomment the line below to include table.tex in this
%% sample file. (Note that you will need to comment out the \documentclass,
%% \begin{document}, and \end{document} commands from table.tex if you want
%% to include it in this document.)


%% The following command ends your manuscript. LaTeX will ignore any text

\end{document}